\documentclass[conference, 11pt]{IEEEtran}
\IEEEoverridecommandlockouts

\usepackage[top=0.75in, bottom=1.2in, left=1in, right=1in]{geometry}

\usepackage{cite}
\usepackage{amsmath,amssymb,amsfonts}
\usepackage{algorithmic}
\usepackage{graphicx}
\usepackage{textcomp}
\usepackage{xcolor}

 \usepackage[numbers, sort&compress]{natbib}

\usepackage[draft]{hyperref}

\hypersetup{
  pdfauthor   = {},
  pdftitle    = {},
  pdfsubject  = {},
  pdfkeywords = {},
  pdfcreator  = {},
  pdfproducer = {}
}

\begin{document}

\title{Should BBR be the default TCP Congestion Control Protocol?}

\author{
    \IEEEauthorblockN{
        Josue Abreu\IEEEauthorrefmark{1},
        Paul Bergeron\IEEEauthorrefmark{1},
        Sandhya Aneja\IEEEauthorrefmark{1}
    }

    \IEEEauthorblockA{\IEEEauthorrefmark{1}School of Computer Science and Mathematics, Marist University, Poughkeepsie, NY, USA}
        
    \IEEEauthorblockA{Emails: josue.abreu1@marist.edu,
    paul.bergeron1@marist.edu,
    sandhya.aneja@marist.edu}
}


\maketitle

\begin{abstract}
In this research, we investigate the feasibility of adopting the Bottleneck Bandwidth and Round-trip propagation time (BBR) protocol as the default congestion control mechanism for TCP. Our central question is whether BBR, particularly its latest iterations, BBRv2 and BBRv3, can outperform traditional TCP variants such as Reno and Cubic across diverse networking environments. We evaluated performance trade-offs in Internet, data center, Ethernet, wireless, and satellite networks, comparing BBR against protocols including DCTCP, DCQCN, TIMELY, HPCC, Swift, and congestion control schemes designed for low-Earth orbit satellite networks, using both experiments and previous studies. Our findings show that BBR consistently achieves high throughput across all environments, with especially strong performance and fairness in scenarios involving homogeneous BBR flows or high bandwidth Internet paths. Experiments with Google and other websites over a 100~Mbps home network further confirm BBR's superior performance and its ability to co-exist with Cubic flows. In another experiment on the Marist campus (1--10~Gbps network), we observed its latency characteristics compared to Cubic. Moreover, a controlled evaluation between protocols reveals that BBR achieves the highest throughput ($\approx 905$~Mbps) but introduces higher latency ($\approx 0.79$~ms) and jitter ($\approx 4.2$~ms). In contrast, Reno and Cubic deliver balanced performance with lower latency and moderate jitter. Vegas prioritizes minimal latency and jitter at the cost of reduced throughput. These results demonstrate the strength of BBR to handle bulk transfers and bandwidth-intensive applications. However, they also emphasize the significance of workload-driven protocol selection in latency-sensitive environments.

\end{abstract}


\section{Introduction}
Linux servers rely heavily on TCP congestion control algorithms to balance throughput and fairness across diverse network conditions. Traditionally, TCP Reno and Cubic have dominated this space, with Cubic emerging as the default in most Linux kernels due to its strong performance in high-bandwidth, high-delay environments \cite{ha2008cubic}. However, with evolving requirements, particularly in latency-sensitive and high-capacity scenarios, there is growing interest in more adaptive and intelligent congestion control protocols. Google's BBR presents such an effort, offering a model-based, RTT-independent approach that seeks to maximize bottleneck bandwidth utilization without persistently filling queues \cite{cardwell2019bbrv2}.

BBRv1 showed promise by operating at Kleinrock's optimal operating point; however, its performance in dynamic environments, such as WiFi and high-mobility networks, revealed limitations \cite{zeynali2024bbrv3}. In response, BBRv2 and BBRv3 were introduced to improve fairness and RTT responsiveness. As Linux increasingly powers systems across cloud, edge, and enterprise contexts, the question becomes whether BBR's modern capabilities justify its adoption as the default protocol across all settings. 

To structure our evaluation, we establish a set of criteria critical for any default congestion control algorithm (CCA) in a general-purpose operating system like Linux:

\begin{enumerate}

    \item{Throughput}:  A default algorithm must achieve high goodput (useful throughput) across a wide range of bottleneck bandwidths. For example, efficiently downloading large files or streaming high-resolution video without interruptions relies on maximizing throughput.

     \item{Latency}: Low and consistent round-trip time (RTT) is critical for user-perceived performance in web browsing, gaming, and video conferencing. An ideal algorithm minimizes queueing delay and reacts quickly to congestion signals without introducing large delay variations (jitter).

      \item{Fairness}: The algorithm must share network capacity fairly with other flows, both those using the same algorithm (intra-protocol fairness) and, crucially, those using different ones, like Cubic (RTT-fairness and inter-protocol fairness). A default CCA should not be overly aggressive. For example, a new algorithm should not starve existing Cubic connections on a shared network link, ensuring a fair experience for all users.

    \item{Robustness}: A default CCA must be robust against non-congestive packet loss (e.g., from wireless errors), sudden changes in available bandwidth, and route changes. Its performance should not degrade in adverse conditions. For example, maintaining a stable video call while moving through an area with fluctuating WiFi signal strength is a key test of robustness.

    \item{Deployment Practicality}: The algorithm should function effectively without requiring universal upgrades to network hardware (e.g., relying on specific Explicit Congestion Notification (ECN) or Active Queue Management (AQM) support that is not widely deployed). For example, an algorithm must work on today's Internet without needing customers to replace their existing home routers.

\end{enumerate}
    The decision to change the Linux default is a complex process that involves the broader Linux kernel community, informed by both academic and industrial research. While no single body holds absolute authority, performance evidence across these criteria, gathered from studies like this survey, is the primary driver for consensus.

Our contribution lies in providing a synthesized analysis of BBR's viability across these criteria. It is essential to emphasize that the conclusions presented are based on our experimental settings and the numbers reported in the studies we cite.

\section{ Types of Congestion Control}
Congestion control methods are generally classified into four types: (i) loss-based, (ii) network-assisted, (iii) delay-based, and (iv) application layer-based approaches. Loss-based mechanisms, such as TCP Reno and Cubic, rely on packet loss as a congestion signal, adjusting the congestion window accordingly. Reno uses additive increase and multiplicative decrease (AIMD), while Cubic improves on Reno by making the window growth more aggressive in long-distance networks \cite{ha2008cubic}. 

Network-assisted mechanisms, such as Quantized Congestion Notification (QCN) and its extensions (e.g., Data Center Quantized Congestion Notification, or DCQCN), utilize explicit feedback from switches to proactively avoid packet loss. Switches send ECN  directly to the sender or via the receiver by marking packets passing through them. When marked packets are sent via the receiver, the scheme employs AQM techniques, such as DCTCP and TIMELY. While in other cases, it is an explicit congestion control protocol, such as ABC and XCP. These are particularly valuable in data centers and lossless Ethernet environments, where performance isolation and fairness are essential. DCQCN, for example, adjusts sending rates based on ECN marks and reduces rate step-wise \cite{chen2025marlin}. However, these methods can suffer from slow convergence and reactive behavior under burst loads \cite{zhang2025ack}. ABC and Cupid, for instance, enhance throughput in wireless networks by dynamically adjusting sending rates based on airtime utilization and active user estimates \cite{du2024revisiting, goyal2020abc}.

Delay-based algorithms, such as TCP Vegas, TIMELY, and BBR Swift \cite{kumar2020swift}, infer congestion from increasing round-trip time (RTTs) rather than losses. While TIMELY and BBR Swift promise lower queuing delays and better responsiveness, they can be unstable in the presence of aggressive loss-based flows in wireless networks. Application-layer control is also used, especially in QUIC-enabled stacks, but remains limited by implementation heterogeneity. 

\section{Internet}
In high-bandwidth Internet settings, Cubic and Reno remain widely used due to their robust implementation and compatibility with legacy infrastructure. However, they often lead to persistent queue buildup and high latency under large-scale traffic loads \cite{ha2008cubic}. Their current versions are New Reno and Cubic with HyStart++.

New Reno remains in a fast recovery state until all lost packets are recovered. New Reno decreases congestion widow (CWND) once within one round-trip time (RTT) when acknowledgments for all lost packets in that RTT are received. Cubic with HyStart++ depends on packet intervals and RTT (delay) information to determine the appropriate time to exit the Slow Start phase \cite{li2024qcc}. TCP Vegas, while offering better latency control, has shown fairness issues and RTT instability (due to noise) when coexisting with loss-based flows. 
Cubic with HyStart++ addressed buffer bloat and thereby latency issues \cite{li2024qcc}. Cubic with New Fast Recovery improved the goodput in environments with non-trivial packet loss. 
These updates advocate for continuing Cubic as a default congestion control. 

BBRv1 introduced a fundamentally different model by avoiding queue buildup altogether and periodically probing for bandwidth. While it demonstrated significant throughput gains in long-distance paths, it also exhibited unfairness toward Reno/Cubic and erratic behavior in RTT-diverse flows \cite{mishra2022we}. BBRv2 mitigated this by introducing a dynamic probing mechanism and more responsive pacing gains, improving convergence fairness in public internet deployments \cite{cardwell2019bbrv2}. 

Recent benchmarks demonstrate that BBRv3 can outperform Cubic in many real-world Internet topologies with high bandwidth-delay product (BDP) paths and shallow buffers by up to 30\% in throughput while maintaining lower queuing delay \cite{zeynali2024bbrv3}. However, Cubic's widespread adoption and stability across low-BDP or fluctuating links suggest that BBR's default status should be contingent on specific application needs. For latency-sensitive, bandwidth-hungry applications such as video conferencing or gaming, BBR is likely the preferable choice. In contrast, fairness-critical scenarios still benefit from Cubic's conservative behavior.

\section{Data Centers}
Congestion control in data centers prioritizes low latency, high throughput, and burst tolerance. Traditional TCP variants, such as Cubic, underperform here due to large queue buildup and incast issues. DCTCP leverages ECN for finer congestion feedback, reducing queue occupancy and latency. However, DCTCP trades off convergence time, disadvantaging short flows that arrive later \cite{alizadeh2010data}. BBRv2 and its variants incorporate ECN support similar to DCTCP and loss-response mechanisms inspired by Cubic. Compared to DCTCP, BBRv2 achieves fewer retransmissions, as well as lower average and 95th percentile RTTs for workloads with 160 or more concurrent flows and micro-burst traffic patterns \cite{cardwell2019bbrv2}.

To address these fairness issues, DCTCP-FQ uses selective marking based on per-flow rate estimation to improve bandwidth sharing. It achieves up to 64.82\% faster convergence time for short flows compared to DCTCP \cite{wangdctcp}. Similarly, BBQ dynamically adjusts ECN thresholds based on buffer conditions, reducing delay while preserving high utilization \cite{zhang2024bbq}. While BBR avoids queue buildup entirely, its aggressive probing can harm latency-sensitive flows during bandwidth transitions. 

Although BBRv3 introduces better convergence handling and reduced queue oscillations, its benefits in tightly coupled, bursty environments, such as data centers, are mixed. However, for the typical mix of short and long flows on 10-100 Gbps links with microsecond-scale RTTs, ACK-driven schemes such as ACC \cite{zhang2025ack} have shown RTT convergence within 50 microseconds.
—far superior to BBR's milliseconds-long response time. 

Therefore, while BBR may complement delay-sensitive protocols, DCTCP-based algorithms remain preferable for intra-datacenter traffic. 

\section{Ethernet}
Lossless Ethernet, particularly in RDMA-over-Converged Ethernet (RoCE) deployments, presents stringent latency and loss constraints. Protocols like TIMELY and DCQCN use RTT or ECN signals to regulate rate while minimizing queue buildup. DCQCN, in particular, remains a standard due to its support for flow isolation and compatibility with PFC\cite{wangdctcp}. 

Nonetheless, DCQCN's slow convergence—typically around 18 milliseconds—has prompted interest in faster alternatives. The ACK-driven congestion control (ACC) method addresses this by improving acknowledgment granularity and flow ramp-up, converging in just 50 microseconds \cite{zhang2025ack}. TCP-based schemes, even BBR, are often unsuitable for lossless environments without substantial adaptation. 

While BBRv2 and v3 are ECN-aware and queue-avoiding, their lack of native support for PFC or coordinated pausing renders them insufficient in lossless Ethernet deployments. Instead, hybrid models such as Swift, which combine DCTCP-style ECN sensitivity with fast ramp-up (as seen in HPCC or ACC), remain more effective. Swift further distinguishes delays incurred at network switches from those at the sender using a cross-layer approach. It incorporates queue offloading at the sender's NIC to reduce end-to-end latency  \cite{kumar2020swift}. Thus, for Ethernet networks, BBR is not a viable default unless significant modifications are made. 

\vspace{5mm}
\begin{figure*}[htbp]

	\centering
	\includegraphics[width=0.7\linewidth]{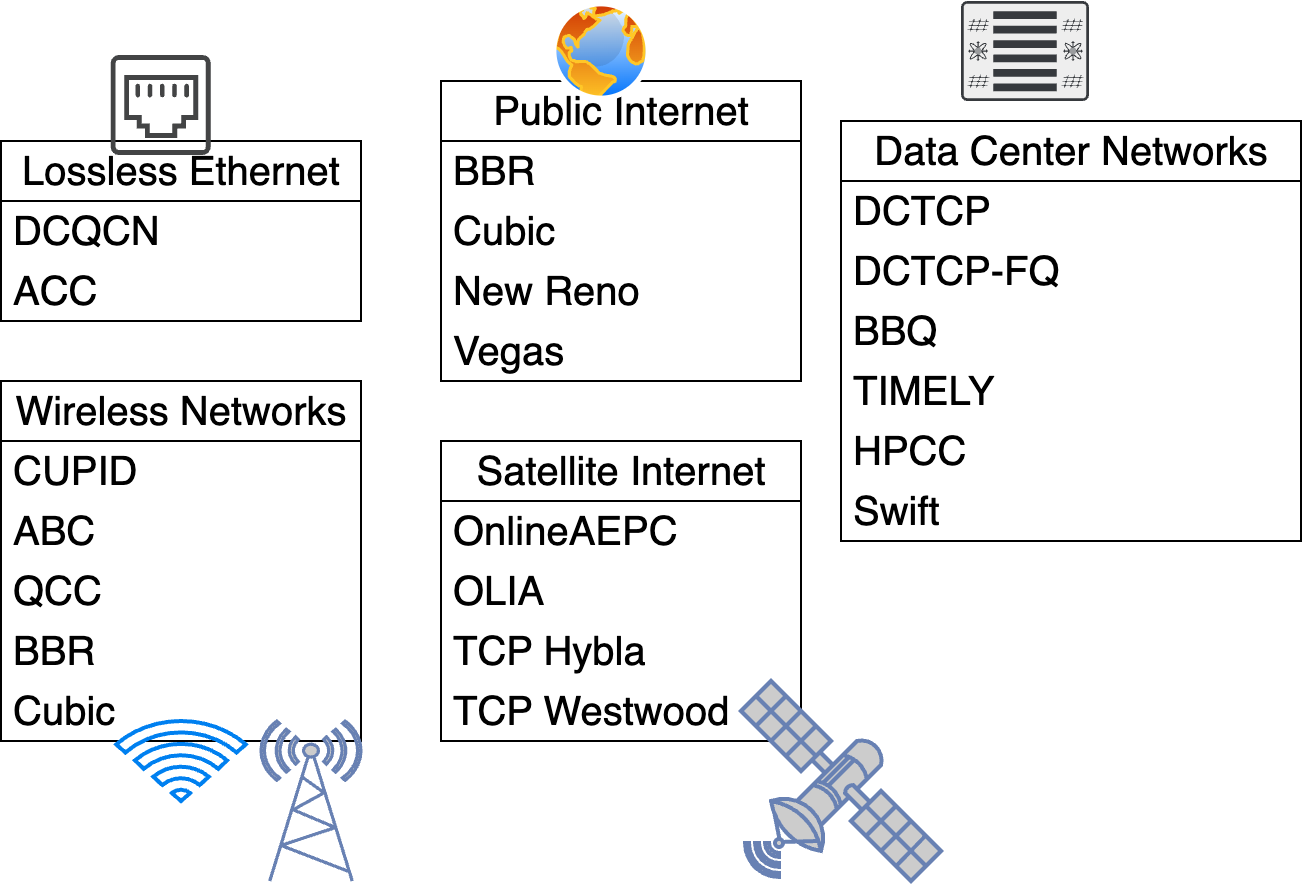}
	\caption{Congestion Control protocols under typical end-host conditions}
	\label{fig:bbr_vs_cubic}
\end{figure*}

\section{Wireless Networks}
Wireless networks suffer from high RTT variance and unpredictable link quality. Traditional TCP variants, such as Reno and Cubic, struggle here due to their inability to distinguish between congestion and wireless-induced losses. Delay-based protocols, such as Vegas and ABC, attempt to address this by incorporating RTT signals and airtime fairness \cite{goyal2020abc}. 

BBRv1 and v2 performed poorly in WiFi networks (e.g., IEEE 802.11ac/n networks with moderate to high cross-traffic) due to their insensitivity to short-term RTT fluctuations, leading to buffer overflow and aggressive retransmissions \cite{du2024revisiting}. BBRv3 introduced better pacing adaptations and conservative gains, improving stability. However, Cupid outperforms BBR in scenarios with multiple mobile users by computing airtime utilization and adjusting rates multiplicatively during congestion \cite{du2024revisiting}. 

QCC \cite{li2024qcc} designed a congestion control mechanism that adjusts the cwnd based on the residual queue length at the sender's NIC after each transmission during file upload. Due to its cross-layered approach, QCC distinguishes between packet losses caused by buffer bloat and link quality variations. QCC reduces delay by 2.36× compared to BBR, 98.5\% of the throughput of BBR, and 3.3x throughput of Copa \cite{ li2024qcc} in WiFi. In Cellular, the scheme achieves 1.39x higher throughput than Sprout \cite{li2024qcc} and 1.81x lower delay than BBR.

Hence, while BBR is less suitable for mobile or shared-channel scenarios, such as 4G/5G or WiFi, it can still serve as a default in static wireless environments with reliable RTT baselines. However, more dynamic congestion controllers, such as ABC, Cupid, or QCC, offer better performance under user mobility and airtime contention. 

\section{Satellite Networks}
Satellite networks present unique challenges due to their high propagation delay, periodic topological changes, and potential for high bit error rates (BER). Traditional congestion control algorithms often overreact to losses, mistakenly identifying them as congestion. TCP Hybla and TCP Westwood were designed to handle such delays but lack modern scalability \cite{fang2024adaptive}.

\citet{fang2024adaptive} introduced OnlineAEPC, a congestion control algorithm that integrates ARIMA-based predictive modeling with adaptive congestion windows. Unlike BBR, which is reactive and model-driven, OnlineAEPC leverages real-time throughput and error prediction. It achieves up to 40.49\% faster transmission at 0.05\% BER compared to Hybla, while reducing packet loss by over 90\%. Yang et al. \cite{yang2024mobility} introduce periodic topology changes and mobility-aware congestion control with Quick start for satellite networks. It outperforms OLIA \cite{yang2024mobility}, a loss-based congestion control, and BBR for GEO bottleneck links. GEO bottleneck links typically involve long RTTs ($\sim$ 500ms), where multipath-aware algorithms using shorter paths often beat traditional loss-based or BBR algorithms.

BBRv3, with its pacing and minimal queuing model, can help avoid retransmissions in noisy environments. However, it lacks native BER handling or predictive congestion windows. As satellite networks continue to grow, especially with LEO constellations, protocols like OnlineAEPC may offer better adaptability than BBR unless BBR evolves to incorporate error correction, multipath \& mobility awareness, and prediction models. 

Figure \mbox{\ref{fig:bbr_vs_cubic}} demonstrates congestion control protocols proposed for bottleneck links under typical end-host conditions. For example, BBR, Cubic, New Reno, and Vegas are used at end devices. ABC is proposed at routers with a few modifications in TCP congestion control. TIMELY, HPCC, DCTCP, DCQCN, DCTCP-FQ, Swift, and BBQ are designed for switches in data centers, with a few modifications to TCP congestion control. BBR Swift also works on the device. CUPID, onlineAPEC, OLIA, and QCC intend to replace classic TCP congestion control specifically for wireless and satellite links. 

\section{Experiments}
We conducted two experiments to evaluate BBR's performance. In the first experiment, we deployed a server in the Hancock ECRL data center at Marist University. Over 15 days, we performed three daily downloads of a 300 MB image across the campus network to compare BBRv1 against Cubic, Reno, and Vegas. The network path in this setup was constrained by 1 Gbps bottleneck link.  In the second, we measured the performance of the top 500 websites over a home network with a 100 Mbps WiFi bottleneck link. Particular attention was given to Google services, such as \textit{maps.google.com} and \textit{scholar.google.com}, among others, since it is widely known that Google utilizes BBR at its data centers.

\section{Results} 

\begin{table}[!tbp]
\centering
\caption{Comparison of congestion control protocols in terms of throughput, latency, and jitter.}
\label{tab:protocol_comparison}
\begin{tabular}{lccc}
\hline
\textbf{Protocol} & \textbf{Throughput (Mbps)} & \textbf{Latency (ms)} & \textbf{Jitter (ms)} \\
\hline
BBR    & 904.61 & 0.79  & 4.21 \\
CUBIC  & 859.94 & 0.04  & 0.28 \\
Reno   & 869.80 & 0.03  & 0.14 \\
Vegas  & 454.59 & 0.02  & 0.12 \\
\hline
\end{tabular}
\end{table}

In the first experiment, the one download of BBR was completed in an average of 25 RTTs, Cubic in 28, Reno in 28, while Vegas in 50 RTTs. There was a notable imbalance in samples from protocols — TCP Vegas 38.6\% of the samples, while the other protocols range between 19.5\% and 21.6\%. The comparative evaluation of congestion control protocols in Table \ref{tab:protocol_comparison} highlights the distinct trade-offs between throughput, latency, and jitter. Among the tested protocols, BBR achieved the highest average throughput at approximately 905~Mbps, clearly outperforming Cubic (860~Mbps), Reno (870~Mbps), and Vegas (455~Mbps). This result reflects BBR's bandwidth-probing design, which actively seeks to maximize utilization of available network capacity. In contrast, Vegas, being primarily delay-based, adopts a more conservative transmission strategy, resulting in significantly lower throughput. Cubic and Reno provide intermediate performance, sustaining high throughput but falling short of BBR's aggressive rate adaptation.  

The latency and jitter characteristics, however, reveal an inverse trend. Vegas demonstrated the lowest average round-trip time ($\approx 0.022$~ms) and minimal jitter ($\approx 0.12$~ms), followed closely by Reno ($\approx 0.031$~ms, 0.14~ms) and Cubic ($\approx 0.036$~ms, 0.28~ms). BBR, while excellent in throughput, exhibited considerably higher latency ($\approx 0.79$~ms) and the largest jitter ($\approx 4.2$~ms). This indicates that BBR’s strategy of maintaining full pipes results in increased delay variability, which can be detrimental for real-time or latency-sensitive applications such as interactive video or gaming. Taken together, these findings suggest that protocol choice should be guided by workload requirements.  BBR is preferable for bulk transfers and high-throughput applications, while Reno, Cubic, or Vegas may be more suitable where latency stability is critical. 

In the second experiment, we considered 78 Google domains in the top 500 websites, and the results show that Google's domains, which are assumed to be using BBR, present a performance profile that differs notably from other websites. Although the aggregate statistics indicate lower average throughput for Google sites ($\approx 0.24$) compared to non-Google domains ($\approx 0.99$), a closer inspection at the domain level reveals a more nuanced picture. Several Google services, such as \textit{insights.sustainability.google} (1.44), \textit{fonts.google.com} (0.60), and \textit{blog.google} (0.59), achieve throughput levels comparable to or exceeding many non-Google websites, all while maintaining low round-trip times in the range of 5--7 ms. This suggests that for content-heavy and sustained flows, Google's implementation of BBR enables the efficient utilization of available bandwidth while minimizing excessive queuing delays.  

At the same time, a large portion of Google's domains demonstrates a prioritization of ultra-low latency over raw throughput. For example, \textit{mail.google.com} achieves a throughput of only 0.016 but maintains an RTT of $\approx 1.5$ ms, while \textit{beinternetawesome.withgoogle.com} records an RTT as low as 0.22 ms. These findings highlight a service-level trade-off: interactive and latency-sensitive applications are optimized for responsiveness, while throughput-intensive services can still exploit BBR's bandwidth-probing capabilities. The result is a deployment strategy where Google ensures consistent low latency across its ecosystem while selectively allowing high throughput when application demands warrant it, demonstrating BBR's adaptability to heterogeneous workloads.  
 
While BBR is inherently bandwidth-aggressive, it can coexist with competing protocols. Earlier studies suggested that BBR is not fair when sharing bandwidth with Cubic \cite{zeynali2024bbrv3}, raising concerns about its suitability as a default protocol. However, Google’s work on BBRv2 shows that mechanisms such as dynamic probing and more responsive pacing gains can improve convergence fairness \cite{cardwell2019bbrv2}. At the same time, independent evaluations caution that intra-protocol fairness issues such as RTT bias and latecomer advantage still persist~\cite{zeynali2024bbrv3}. Importantly, these fairness outcomes appear to be highly \textit{environment-dependent}: in well-managed networks with ECN support or shallow buffers, BBRv3 often coexists equitably with Cubic, whereas in heterogeneous Internet conditions with deep buffers, variable RTTs, or staggered flows, fairness deteriorates significantly. Our experiments on the top 500 websites demonstrate that BBR coexists with BBR, Cubic, QUIC, and other widely deployed congestion control algorithms across the Internet.

\section{Conclusion}

In summary, BBR's innovative congestion control model makes it an attractive candidate for adoption as the Linux default in scenarios characterized by long-lived flows, high-bandwidth paths, and environments that require RTT stability. Our results show that BBR sustains the highest throughput ($\approx 905$~Mbps) compared to Reno, Cubic, and Vegas, confirming its strength in maximizing bandwidth utilization. However, this performance comes at the cost of higher latency ($\approx 0.79$~ms) and greater jitter ($\approx 4.2$~ms), in contrast to Reno, Cubic, and especially Vegas, which prioritize latency stability. This trade-off highlights that while BBR is well-suited for bulk transfers and bandwidth-intensive applications, selecting a protocol in latency-sensitive environments requires careful consideration. The survey also highlights important limitations of BBR in data centers, lossless Ethernet, wireless, and satellite networks.  Nonetheless, the ongoing evolution of BBR has addressed many of its earlier shortcomings.

\bibliographystyle{IEEEtranN}

\bibliography{refs}

\bibliographystyle{IEEEtranN}
\bibliography{refs}

\end{document}